\documentclass[a4paper,UKenglish,cleveref, autoref, thm-restate]{lipics-v2021}

\title{The effect of disruptive events on spatial and social interactions: An assessment of structural changes in pre-and post-COVID-19 pandemic networks} 

\titlerunning{The effect of disruptive events on spatial and social interaction networks} 

\author{Caglar Koylu\footnote{Corresponding author}}{Department of Geographical and Sustainability Sciences, The University of Iowa, USA \and \url{https://geo-social.com} }{caglar-koylu@uiowa.edu}{https://orcid.org/0000-0001-6619-6366}{This work is in part supported by The National Science Foundation (NSF) Grant No. 2215568 titled “Population-scale kinship networks and migration”}

\author{Maryam Torkashvand}{Department of Geographical and Sustainability Sciences, The University of Iowa, USA}{maryam-torkashvand@uiowa.edu}{https://orcid.org/0000-0003-4318-8085}{}

\authorrunning{C. Koylu and M. Torkashvand} 

\Copyright{Caglar Koylu and Maryam Torkashvand} 



\category{} 

\relatedversion{} 

\nolinenumbers 

\EventEditors{J. Long, S. Dodge, U. Demsar, R. Weibel}
\EventNoEds{4}
\EventLongTitle{GIScience 2023 Workshop on Disruptive Movement Analysis (DMA)}
\EventShortTitle{GIScience 2023}
\EventAcronym{GIScience 2023 DMA’23}
\EventYear{2023}
\EventDate{September 12, 2023}
\EventLocation{Leeds, United Kingdom}
\EventLogo{}
\SeriesVolume{}
\ArticleNo{1}

\begin{document}

\maketitle

\begin{abstract}
Disruptive events significantly alter spatial and social interactions among people and places. To examine the structural changes in spatial and social interaction networks in pre- and post-periods of the COVID-19 pandemic, we employ the Louvain method to algorithmically detect regions (communities) within the county-to-county networks of the SafeGraph mobility and Facebook social connectedness. We then utilize a range of partition similarity metrics, including adjusted Rand, z-Rand, Normalized Mutual Information (NMI), and Jaccard indices, to quantitatively measure the similarity of regions between the pre- and post-periods partitions of each network. Our findings reveal that in the post-pandemic period, spatial interactions led to the formation of localized geographic communities or regions characterized by higher modular activity within each region. In contrast, online social interactions shifted towards longer distance connections, resulting in the emergence of larger regions marked by strong friendship ties that often encompassed multiple states. By understanding these changes, we contribute to a better comprehension of the pandemic's impact on our interconnected physical-virtual world, providing valuable insights for future research and informing strategies to adapt to the evolving dynamics of human interactions.

\end{abstract}

\section{Introduction}
\label{sec:typesetting-summary}

Disruptive events substantially change the spatial and social interactions of both humans and animals. While existing studies have extensively examined the volumetric, temporal and spatial impact of such events on human activities, the changes in structural patterns of human movement and communication networks remain largely unexplored. For example, the COVID-19 pandemic drastically changed how people move, communicate and form connections in real-world and virtual settings. The pandemic has resulted in a reduction in the frequency of human movement, leading to more localized and sparser interactions in physical spaces \cite{schlosser2020covid, gao2020mapping}. Simultaneously, online interactions have become denser as individuals spent more time at home, shifting their social interactions from the physical world to the virtual realm due to lockdown measures \cite{pandya2021social}. Despite these observed trends, the implementation of a national and uniform set of non-pharmaceutical interventions (NPIs) has proven to be ineffective \cite{baghersad2023modularity}. This is because both the incidence of the disease and human interactions exhibit spatiotemporal heterogeneity. The impact of the pandemic and the effectiveness of interventions vary across different geographical locations and evolve over time. Considering the spatial and temporal variations, only a limited number of studies implemented community-detection algorithms to construct geographic regions that could accurately reflect natural human movement and relationships \cite{andris2021human, buchel2021strategizing, gibbs2021detecting}. These constructed regions can serve as a foundation for implementing targeted containment measures at a regional level, thereby enabling more effective disease control strategies. However, despite these endeavors, there remains a lack of systematic comparative analysis regarding the changes in structural patterns (regions) within spatial networks before and after a disruptive event.

In this study, we introduce a network comparison workflow to assess the structural changes in spatial and social networks during the pre- and post-periods of the COVID-19 pandemic, utilizing two datasets: the SafeGraph mobility \cite{safegraph2020} and Facebook's Social Connectedness Index \cite{bailey2018social}. Specifically, we derive county-to-county SafeGraph mobility data spanning from March 2019 to March 2020, representing the pre-pandemic period, and from April 2020 to April 2021, representing the post-pandemic period. SafeGraph mobility data is originally at census block groups such that each flow is from one block group to another. We simply aggregate block group-level flow data to county-level flow data. To explore the evolving structure of social interactions in virtual space, we use the Facebook Social Connectedness Index (SCI) at the county level in 2015, representing the pre-pandemic, and in 2021, representing the post-pandemic period. SCI represents the number of Facebook friends between user accounts in two counties divided by the product of the numbers of accounts in those counties.

\section{Methodology}

There are two main methods for comparing networks: Unknown Node-Correspondence (UNC) and Known Node-Correspondence (KNC) \cite{tantardini2019comparing}. UNC methods allow for the comparison of any pair of networks, even if they differ in size, density, or domain. In contrast, KNC methods specifically compare two networks where the pairwise correspondence between nodes is known. This means that the two networks being compared using KNC should either have the same set of nodes or at least a common subset of nodes. For our analysis, we utilize a KNC workflow to examine the changes in community structures in pre- and post-event networks.

To identify the community structures (regions), we first employ the Louvain algorithm \cite{blondel2008fast} on the county-to-county Facebook and SafeGraph networks of both pre- and post-pandemic periods across the continental US. The Louvain algorithm evaluates the strength of the partitioning of a network into modules using a measure of modularity. A network with dense connections between nodes within modules but sparse connections between nodes in different modules generates high modularity. The Louvain algorithm does not enforce spatial contiguity, thus, allows communities to be geographically distant and disjoint from one another. 

Although the similarity of communities between two partitions could be revealed visually, it is essential to quantitatively compare the agreement between the detected communities. We use a series of partition (or cluster) similarity metrics including adjusted Rand, z-Rand, Normalized Mutual Information (NMI), and Jaccard indices to quantify the similarity of the community structures (regions) between partitions of pre- and post-periods of each network. The most common measure is the Rand coefficient, which is the count of pairs that clustered in the same way (either in the same or different partitions across both networks) over the total pairs \cite{rand1971objective}. The Rand coefficient is biased toward random partitions. To address this issue, the Adjusted Rand Index subtracts the expected value due to randomly constructed partitions and normalizes the result regarding the maximum threshold and mean value \cite{hubert1985comparing}. In other words, the Adjusted Rand index method mitigates high values for two randomly related partitions using a maximum bound. The z-Rand or z-score of the Rand coefficient \cite{red2011comparing} computes the number of node pairs that belong to the same community in two different partitions. The z-Rand score provides statistical inference by comparing the pair-count captured by the Rand coefficient to its expected value under a null model with the same size communities. The Normalized Mutual Information (NMI) is an entropy measure that evaluates the amount of shared information on two partitions \cite{estevez2009normalized}. A value of 0 denotes that there is no mutual information between partitions and that two partitions are completely different, while a value of 1 shows identical partitions. Jaccard coefficient assesses the extent of overlap between two partitions by calculating the ratio of the intersection of the partitions to their union. Similar to NMI, a Jaccard coefficient of 0 indicates no overlap or similarity between two partitions, whereas a value of 1 means a perfect match between two partitions. 

To gain insights into the geographical characteristics of regions during the pre- and post-periods, we compute two spatial metrics. First, we calculate the average border length per region by dividing the shared border length between regions by the number of regions. Second, we compute compactness \cite{polsby1991third} of each region in each partition. We then compute the mean and median compactness measures for each partition, which offer insights into the change in the overall shape and dispersion of regions.

\section{Results}

We algorithmically detect the regions (communities) from county-to-county networks of Safegraph mobility and Facebook social connectedness for pre- and post-pandemic periods using the Louvain method of modularity maximization \cite{blondel2008fast}. The hypothesis that online social connections were intensified after the pandemic can be observed by the comparison of region partitions of pre- and post-Facebook connectedness networks. Modularity significantly decreased from 0.87 in pre-period to 0.82 of the post-period partition as a result of smaller number of regions with larger areas in the post-pandemic period. Most notably, the West expanded to contain parts of Utah, and Arizona whereas Mid-Atlantic region in the East expanded to include Carolinas, Virginia, Maryland and Southern Pennsylvania. From many smaller regions across the country reconciled into larger regions, which include but are not limited to: (1) Wisconsin, Minnesota, Northern Chicago, and Iowa; (2) Illinois and Indiana; (3) Louisiana, Mississippi, and Alabama (4) West and East Texas; (5) Kansas and Colorado; (6) Nebraska and Wyoming.

On the other hand, the hypothesis about the decrease in non-essential and long-distance travels due to lockdowns are also observed by the comparison of pre- and post-period partitions derived from SafeGraph data. Decrease in long-distance flows resulted in a slight increase in the number of regions from 53 in pre- to 57 in the post-period with also a slight increase in modularity. Most notably, regions that approximately matched state borders in the pre-period such as Georgia, South Carolina, Illinois, and Tennessee shrunk in size by splitting into multiple small regions. Meanwhile some regions increased in size such as North Carolina taking northern counties of South Carolina. 

When comparing the regions of online social connections and the human mobility, there are interesting differences. For example, Upper Peninsula of Michigan (north of Wisconsin) remain to be in the same region with the mainland Michigan in both pre- and post-pandemic regions of Facebook. In contrast, the western part of Upper Peninsula resides within Wisconsin region for both pre- and post-pandemic mobility networks. This suggests that Upper Peninsula is virtually more connected to Michigan (stronger friendships with mainland Michigan), whereas it is more physically connected to Wisconsin (stronger mobility flows or spatial interactions with Wisconsin). 

For each partition, we compute the average shared border length per region by dividing the total length of borders among regions by the number of regions. In the pre-period, the network of Facebook connectedness consisted of 32 regions, with an average border length of 1055 km. In the post-period, there were 25 regions with an average border length of 1057 km. Although the total shared border length increases in partitions with larger number of regions, the average border length remained similar in Facebook partitions possibly due to the large western regions in the post-period partition. On the other hand, SafeGraph mobility regions exhibit a reduction of 37 Km in the post-period partition mainly because the increase in the number of regions are due to the addition of small regions in the East, which did not affect border lengths significantly.

Additionally, we examined the mean and median measures of compactness within each network. A noticeable decrease in compactness values is observed for the Facebook regions from the pre-pandemic to the post-pandemic period. Conversely, the SafeGraph mobility regions exhibit consistent compactness values, even as the number of regions increases from 53 to 57. 

\begin{figure}[h]
\centering
\includegraphics[width=1\linewidth]{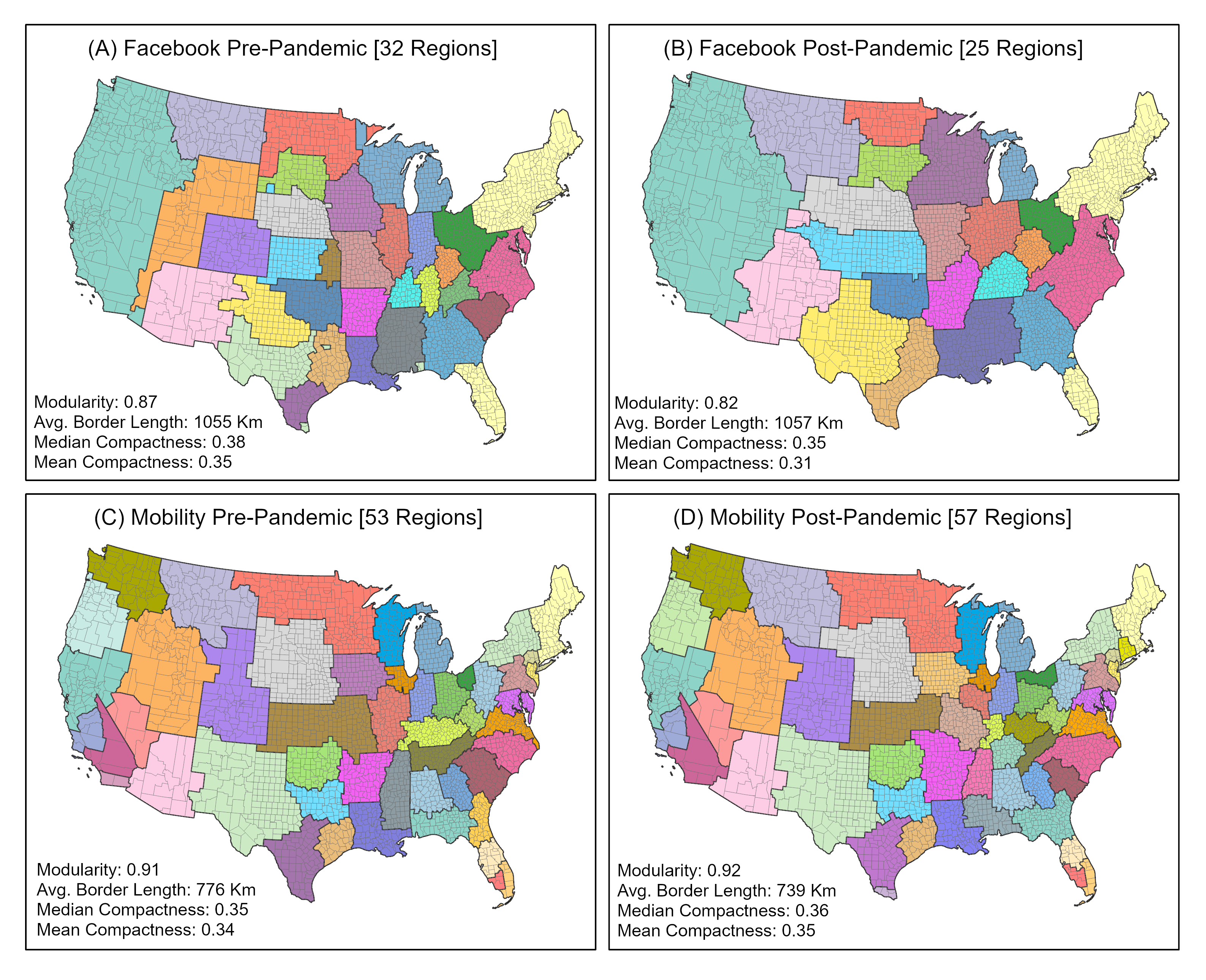}
\caption{We apply the Louvain algorithm to detect regions for the county-to-county networks of (A) Facebook Social Connectedness Index of 2015 (Pre-Pandemic) (B) Facebook Social Connectedness index of 2021 (Post-Pandemic) (C) SafeGraph Mobility: March 2019 - March 2020 (Pre-Pandemic) (D) SafeGraph Mobility: April 2020 - April 2021 (Post-Pandemic) 
}
\label{fig:schematics}
\end{figure} 

Table 1 presents partition similarities of the pre- and post-pandemic Facebook and SafeGraph networks using various metrics: z-scores of the Rand coefficient, Adjusted Rand index, NMI, and Jaccard coefficient. Although each of these measures reveal different aspects of the overlap between pre- and post-period partitions, the SafeGraph mobility networks exhibit greater regional similarity compared to the Facebook networks across all measures. The high z-Rand scores reveal that the SafeGraph and Facebook regions of pre- and post-periods highlight significant correlations between the post and pre networks than should be expected at random. The Facebook regions produce 0.67 and 0.52 for the Adjusted Rand index and Jaccard index, respectively, indicating a slightly higher level of change as compared to the Adjusted Rand index of 0.82 and Jaccard index of 0.70 for the SafeGraph regions. These trends are further supported by the NMI index, which indicates higher shared information between regions in the pre- and post-pandemic SafeGraph networks compared to the Facebook networks. Collectively, these structural comparison measures, along with our visual comparison with the maps in Figure 1, confirm both similarities and changes in human mobility and connections in both physical and virtual environments following the pandemic. 


\begin{table}[htb]
\centering
\caption{Partition similarity for pre-and post event networks}
\begin{tabular}{lrrrrrr}
\hline
\multicolumn{1}{c}{Network} &
\begin{tabular}[c]{@{}r@{}}Pre-Period\end{tabular} &
\begin{tabular}[c]{@{}r@{}}Post-Period\end{tabular} &
\begin{tabular}[c]{@{}r@{}}z-Rand\end{tabular} &
\begin{tabular}[c]{@{}r@{}}Adj. Rand\end{tabular} &
\begin{tabular}[c]{@{}r@{}}Jaccard\end{tabular} &
\begin{tabular}[c]{@{}r@{}}NMI\end{tabular}\\ 
\hline
SafeGraph & 2019/3-2020/3 & 2020/4-2021-4 & 1656.75 & 0.82 & 0.70 & 0.93\\
Facebook & 2015 & 2021 & 1105.38 & 0.67 & 0.52 & 0.83\\
\hline
\end{tabular}
\label{tab:nets}
\end{table} 

\section{Conclusion}

This study presents the preliminary results of a comparative evaluation of the structural changes in two spatial and social interaction networks for pre- and post-COVID-19 pandemic periods: the SafeGraph mobility network, and Facebook's social connectedness. Our findings highlight that the effect of the pandemic on online social interactions and spatial interactions are very distinct. While spatial interactions resulted in more localized geographic communities or regions with higher modular activity within regions, online social interactions switched to longer distance connections and thus, larger regions of strong friendship ties that often aggregate multiple states. 

Impacts of disruptive events on human movement and social interactions can vary widely depending on the nature of the event and the context. However, our approach to analyzing the structural changes in spatial and social interaction networks is not unique for studying the COVID-19 pandemic and can be applied to study the effect of disruptive events at larger or smaller geographic and temporal scales in different domains. In future work, we plan to (1) expand our workflow to include more comprehensive set of network comparison methods; and (2) study the effect of another disruptive event in the U.S. history, the Civil War, on the migration networks, which we derive from population-scale family tree \cite{koylu2021connecting,koylu2022measuring} and linked census data \cite{helgertz2022new}. 



\bibliography{lipics-v2021-sample-article}

\end{document}